# Electronic Raman scattering of Tl-2223 and the symmetry of the superconducting gap

A. Hoffmann, P. Lemmens, G. Güntherodt, V. Thomas [*], K. Winzer [*]

2. Physikalisches Institut, RWTH-Aachen, 52056 Aachen, Germany

[*] 1. Physikalisches Institut, Universität Göttingen, 37073 Göttingen, Germany

Single crystalline $Tl_2Ba_2Ca_2Cu_3O_{10}$ was studied using electronic Raman scattering. The renormalization of the scattering continuum was investigated as a function of the scattering geometry to determine the superconducting energy gap $2\Delta(\hat{k})$. The $A_{1g}$- and $B_{2g}$-symmetry component show a linear frequency behavior of the scattering intensity with a peak related to the energy gap, while the $B_{1g}$-symmetry component shows a characteristic behaviour at higher frequencies. The observed frequency dependencies are consistent with a $d_{x^2-y^2}$-wave symmetry of the gap and yield a ratio of $2\Delta_0 / k_B T_c = 7.4$. With the polarization of the scattered and incident light either parallel or perpendicular to the $CuO_2$-planes a strong anisotropy due to the layered structure was detected, which indicates an almost 2 dimensional behavior of this system.

## 1. INTRODUCTION

The symmetry of the energy gap $2\Delta(\hat{k})$ is an important aspect in understanding high $T_C$ superconductivity, because this symmetry may give an important clue to a possibly unconventional pairing mechanism. A pairing mechanism based on spin fluctuations should lead to a $d_{x^2-y^2}$-wave symmetry, while other pairing mechanisms may lead to a different symmetry of the order parameter. In spite of major efforts to determine this symmetry experimentally, this question is still unresolved [1].

Electronic Raman scattering provides information about the symmetry of the amplitude of the energy gap $2\Delta(\hat{k})$, whereas it is not sensitive to a phase shift in the order parameter. Even so, since the ratio of the optical penetration depth to the coherence length is large in high $T_C$ superconductors, electronic Raman scattering probes bulk properties, in contrast to tunnelling or photoemission spectroscopy, and hence gives more reliable information.

In this work $Tl_2Ba_2Ca_2Cu_3O_{10}$ was examined, since it is the only high $T_C$ superconductor with three $CuO_2$ per unit cell available as high quality single crystals. The investigated crystals have the highest $T_C$ ( = 118K) of all materials studied until now by electronic Raman scattering [2]. For the Raman measurements we used the 476.5nm excitation line of an $Ar^+$-laser in quasi-backscattering geometry, with a power level below 15 W/cm². The scattered light was analyzed using a Dilor XY spectrometer, with a spectral resolution of 8.5cm$^{-1}$. The sample was cooled in He contact gas.

Electronic Raman scattering of free carriers occurs due to massfluctuations about the Fermi surface. A continuous scattering background up to high frequencies is observed in high $T_C$ superconductors, whose origin is still unclear. For temperatures below $T_C$ this scattering background becomes renormalized for a frequencies below an average of the energy gap $2\Delta(\hat{k})$ over the Fermi surface.

## 2. S- VS. D-WAVE

Different symmetry components of the electronic Raman scattering $A_{1g}$, $B_{1g}$, and $B_{2g}$ can be measured using polarized light. For each of the symmetry components the amplitude of the order parameter $2\Delta(\hat{k})$ gets a different weight for a particular direction of k, which is important for the case of an anisotropic order parameter. Assuming an order parameter with $d_{x^2-y^2}$-wave symmetry the symmetry components were calculated in a recent theory. For the $B_{1g}$-component a peak at $2\Delta_0$ (with $\Delta_0$ being the maximum value of the gap) is predicted, while the $A_{1g}$- and $B_{2g}$-component should have a peak at lower frequencies. This is indeed observed in $Bi_2Sr_2CaCu_2O_{8+\delta}$ [3]. On the other hand an isotropic s-wave symmetry of the gap should lead to a peak at $2\Delta$ for all symmetry components.

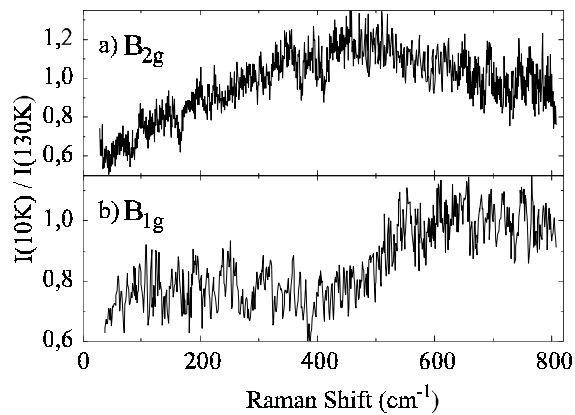

Figure 1. Normalized intensities of the a) $B_{2g}$ and b) $B_{1g}$ symmetry component of Tl-2223.

In order to suppress phonon peaks, and to emphasize the redistribution of the electronic Raman scattering intensity below $T_c$, spectra at T=10K are divided by spectra at T=130K, see fig. 1. The $B_{1g}$-component is suppressed below 610cm$^{-1}$, while the $B_{2g}$-component has a peak at 450cm$^{-1}$.

The normalized $A_{1g}$-component is shown in fig. 2 a). Its frequency behavior is similar to the $B_{2g}$-component with a linear rise at low frequencies and a peak at 430cm$^{-1}$. For all three symmetry components the frequency behavior and the positions of the peaks are consistent with the theory assuming a $d_{x^2-y^2}$-symmetry of the energy gap [3]. The gap ratio is determined as $2\Delta_0 / k_B T_c = 7.4$.

## 3. LAYERED STRUCTURE

In addition we made measurements with the polarization perpendicular to the $CuO_2$ planes. In this case only the $A_{1g}$-symmetry component can be measured, which is shown in fig. 2 b), for comparison with the corresponding measurement with the polarization parallel to the $CuO_2$ planes. This symmetry component is suppressed below 200cm$^{-1}$. By contrast for polarization parallel to the $CuO_2$ planes the same symmetry component is suppressed below 400cm$^{-1}$.

A.A. Abrikosov showed that this factor of 2 is to be expected for the case of an extremely weak coupling between superconducting and normal conducting layers in a layered superconductor [4]. This indicates an almost 2-dimensional behavior of $Tl_2Ba_2Ca_2Cu_3O_{10}$.

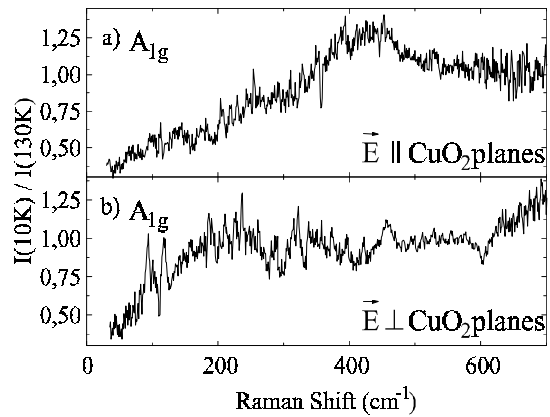

Figure 2: $A_{1g}$-component for polarization a) parallel and b) perpendicular to the $CuO_2$ planes of Tl-2223. The spectrum in a) was obtained by subtracting the $B_{1g}$-spectra from the $A_{1g}+B_{1g}$-spectra.

## 4. CONCLUSION

A complete symmetry analysis of the electronic Raman scattering of $Tl_2Ba_2Ca_2Cu_3O_{10}$ below $T_c$ has been carried out. With the polarization parallel to the $CuO_2$ planes, the different symmetry components show a different frequency behavior for $B_{1g}$ vs. $A_{1g}$ and $B_{2g}$ respectively. These results are similar to measurements on $Bi_2Sr_2CaCu_2O_{8+\delta}$ and $YBa_2Cu_3O_{7-\delta}$ single crystals [2] and are consistent with a theory which takes a $d_{x^2-y^2}$-wave symmetry of the energy gap into account [3]. Furthermore, the spectra taken with the polarization perpendicular to the $CuO_2$ planes sample the weak coupling of the superconducting layers, which is due to the strong anisotropy of the layered structure.

We thank D. Einzel, T.P. Devereaux and M. Cardona for useful discussions. This work was supported by DFG through SFB 341.